# Virtual single-photon transition interrupted: time-gated optical gain and loss

Jens Herrmann[1*], Matthias Weger[1], Reto Locher[1], Mazyar Sabbar[1], Paula Rivière[2,3], Ulf Saalmann[3], Jan-Michael Rost[3], Lukas Gallmann[1], Ursula Keller[1]

[1]Physics Department, ETH Zurich, CH-8093 Zürich, Switzerland.

[2]Departamento de Química, Universidad Autónoma de Madrid, ES-28049 Madrid, Spain.

[3]Max Planck Institute for the Physics of Complex Systems, D-01187 Dresden, Germany.

*Corresponding author. Email: jens.herrmann@phys.ethz.ch

**Introductory paragraph**

The response of matter to an optical excitation consists essentially of absorption and emission. Traditional spectroscopy accesses the frequency-resolved and time-integrated response, while the temporal evolution stays concealed. However, we will demonstrate here that the temporal evolution of a virtual single-photon transition can be mapped out by a second pulsed electromagnetic field. The resulting optical signal shows previously unexpected optical gain and loss, which can be gated and controlled via the relative delay of the electromagnetic fields. The model presented here can be applied to any system that assumes a two-level character through near-resonant optical dipole excitation, whether they are of atomic, molecular or even solid-state nature. These theoretical observations are in excellent qualitative agreement with our transient absorption spectroscopy study in helium[1-3]. The presented results can act as starting point for a new scheme for creating optical gain, which is a prerequisite for the operation of lasers. It may be possible to open the doors to spectral regions, which were difficult to access until now, e.g. in the extreme ultraviolet.

The development of transmission as well as absorption spectroscopy has been crucial for understanding some of nature's secrets. In particular the laser opened the door to a variety of new spectroscopic techniques. One of these techniques, namely femtosecond time-resolved spectroscopy, had a major impact in physics, photonics and chemistry[4]. Most recently time-resolved transient absorption spectroscopy was applied in the attosecond and extreme ultraviolet (XUV) regime for the first time[1-3] benefiting from the rapid progress in attosecond pulse generation[5-10]. Absorption through resonant excitation of bound-bound transitions and direct ionization play a key role in these experiments and recent theoretical work[11-18]. Since matter is essentially transparent to off-resonant electromagnetic radiation this was not considered in previous attosecond transient absorption studies. Regardless, we will show that the off-resonant excitation can be used to achieve optical gain.

We investigate the simplest model for laser-matter interaction, a two-level system (TLS). The knowledge of the temporal evolution of the TLS's dipole response is essential for the complete understanding of light-matter interaction. This evolution is non-observable in common optical experiments, since they only access the time-integrated dipole response (TIDR). Our theoretical time-frequency analysis helps to overcome this experimental limitation (see Fig. 1c and Fig. 1d). In the case of off-resonant excitation the TIDR vanishes, which implies that the material is transparent for the incident, exciting electromagnetic field (see Fig. 1e, solid blue line). Nevertheless, the dipole shows an oscillatory behavior with positive and negative contributions. Hence, the transparency is the result of a delicate balance of these nonzero contributions. The insight into the temporal evolution with the help of our time-frequency analysis allows us to develop a method to access the nonzero contributions by manipulating the evolution before it is finalized. For this purpose we apply a second electromagnetic field, which interrupts the evolution and unbalances positive and negative contributions. The resulting nonzero TIDR can be experimentally accessed and results in optical gain and loss. Furthermore, the amount of loss or gain can be controlled via the relative delay between the two electromagnetic fields. By not relying on any system-specific parameters, our formulation reveals fundamental properties of off-resonant light-matter interaction where the matter part is modeled with a universal TLS. The interaction of two laser fields with an atomic TLS has already been studied in previous investigations[19-21]. Nevertheless, these studies were performed in the monochromatic limit, i.e. using continuous-wave lasers or with a fixed delay between the two laser pulses. They did not have the possibility to vary the relative delay between the electromagnetic fields, which is a crucial parameter in our experiment and theoretical study.

The TLS is the simplest quantized structure that allows exchange of energy with the environment[22]. Ground state $|g\rangle$ and excited state $|e\rangle$ are separated energetically by the transition energy $\Delta$. The system responds to electromagnetic radiation through the transition dipole $d(t)$, given by $d(t) \propto a(t) + a^*(t)$, where $a(t)$ is the amplitude of $|e\rangle$. We excite our system with a light pulse, which is off-resonant with regard to the transition between the two levels and so weak that it can be treated perturbatively. The actual pulse parameters of 23.37 eV center energy and 20 fs duration are in principle arbitrary, but chosen here to match our experiment as further discussed below. Fig. 1c illustrates the dipole response in the time-frequency representation. The response intrinsically comprises positive and negative contributions of comparable magnitude. The dipole in rotating-wave approximation follows the electric field $\widetilde{V}(t)$ of the excitation pulse with angular frequency $\widetilde{\omega}$ (atomic units are used),

$$d(t) = \frac{1}{\widetilde{\omega} - \Delta} \widetilde{V}(t) \qquad (1)$$

for off-resonant excitation, i.e. $|\tilde{\omega}-\Delta|\tilde{T}\gg 1$ with $\tilde{T}$ the pulse duration of the excitation pulse. Only in this case it exhibits the oscillatory pattern of emission and absorption in a time-frequency resolved representation (Fig. 1c). Nevertheless, the pattern vanishes upon temporal integration (Fig. 1e, solid blue line) and is non-accessible with traditional spectroscopy. Additionally, we plot the TIDR when integrated only over negative or positive times in Fig. 1e. This illustrates how the negative- and positive-time contributions cancel each other out.

Now we show how to resolve and control this fundamental process in time by adding a control pulse to modulate the system. This results in real emission and absorption.

Without loss of generality, but motivated by our experiment, we used 1.57 eV photon energy and 30 fs duration for the control pulse. The off-resonant excitation pulse couples weakly to the TLS and can be treated in first-order perturbation theory. The control pulse $\overline{V}(t)$ with an amplitude orders of magnitude larger than the excitation pulse, precluded a perturbative treatment. Hence, the amplitude $a(t)$ was described in perturbation theory only with respect to the excitation pulse $\tilde{V}(t)$ by the time-dependent Schrödinger equation

$$i\frac{\partial}{\partial t}a(t)=[\Delta+\overline{V}(t)-i\gamma\overline{V}_t]a(t)+\tilde{V}(t) \qquad (2)$$

where we added a damping term proportional to the control pulse envelope $\overline{V}_t$. It takes into account the loss of population to other states or the continuum of a real physical system (see Fig. 2b). The intensity of the moderately strong control pulse was chosen such that it was strong enough to periodically modulate the resonance energy (which elsewhere[23] has been referred to as Stark modulation) of the TLS but sufficiently weak to avoid multiphoton excitations. Figure 1d displays the dipole response of the TLS for the off-resonant excitation pulse overlapping the control pulse with a lead of 15 fs (Fig. 1b). For times $t > 0$ the dipole response is significantly suppressed resulting in a nonzero TIDR. A positive response represents absorption, while a negative response corresponds to a net emission of photons (red and blue in the color scale of Fig. 1, respectively). We interpret the net emission as optical gain since the photon count in the affected spectral regions is increased with respect to the spectrum of the incident pulse. Integration over the full spectrum yields no net emission of photons, due to energy conservation.

We expect a similar response for quite diverse systems with appropriate intensity and photon energy scaling since our model does not rely on any system specific property.

The TIDR for different time delays between the two electromagnetic fields can be measured in a pump-probe experiment. Fig. 3a shows the expected behavior. A negative delay corresponds to the control pulse arriving before the excitation pulse. This delay-frequency representation of the TIDR is a fundamentally different quantity compared to the time-frequency analysis of the dipole response in Fig. 1. The simulated pump-probe data shows that for certain pulse delays a strong positive or negative dipole response, i.e. optical loss or gain, remains after the temporal integration. The TIDR is symmetric around 0-fs delay and point symmetric about the center frequency of the excitation pulse. The relative delay between the two fields determines magnitude and sign of the response and we can switch between absorption and gain in a given spectral window.

We tested our model by comparing it with a pump-probe experiment. For the TLS we chose a quantum-mechanical prototype system: helium (He) in its $1s^2$ ground state. The 1s3p state represents the excited state $|e\rangle$. We used an attosecond pulse train (APT) for the off-resonant excitation. The spectrum of the APT is composed of discrete peaks at odd multiples of the generating laser pulse carrier frequency[6]. The 15th harmonic has a central energy of 23.37 eV (FWHM ~130 meV), well below the ionization potential of He (24.59 eV)[24]. The 15th harmonic was off-resonant with respect to the 1s3p and 1s4p states. The use of a single attosecond pulse is not possible in our case. The broad continuous spectrum would simultaneously populate several excited states and prevent an off-resonant excitation. Figure 2b depicts the investigated spectral window, with the relevant states and the ionization threshold. We neglected the 1s4p state, since the dipole matrix element for the transition from the ground state is by a factor of three smaller compared to the 1s3p state[25].

For the control pulse we used a moderately strong IR pulse, which was separated from the driving IR pulse before the generation of the APT (from now on referred to as XUV pulse) and sent on an independent beam path (see Fig. 2a). This path length can be controlled in order to set the timing between the XUV and the IR pulse. This enables us to vary the relative delay between the two pulses, which allows one to control absorption and emission in the *time domain* in a fundamentally different way compared to former investigations[19-21]. An XUV sensitive grating spectrometer detected the transmitted XUV photons after the residual IR light had been filtered out.

Figure 2c displays the transmitted XUV yield for different delays between the two pulses and an IR intensity of $4.2 \times 10^{12}$ W/cm$^2$. The transmitted signal exhibits a strong reduction when the peak of the XUV and IR pulse coincide in the target around ~0-fs delay. Since the 15th harmonic is off-resonant and energetically below the ionization potential of He, absorption only occurred through two-photon absorption with an XUV photon and an assisting IR photon by exciting an electron into the continuum via a virtual dipole transition[26]. To analyze the pump-probe data we plotted the change in absorbance $\Delta\alpha = \ln(\tilde{I}/I)$ as induced by the IR gating pulse in Fig. 3b. Here, $\tilde{I}$ and $I$ represent the harmonic signal without and with the IR pulse, respectively. Positive $\Delta\alpha$ indicates absorption and is displayed in red. The strong decrease of the transmitted signal related to the two-photon absorption via the virtual state is still evident. It is energetically located around the center energy of the 15th harmonic. As predicted in the model, we also observed a net-emission of XUV photons, which corresponds to a negative $\Delta\alpha$ (blue). The strength of the optical gain is of the same order of magnitude as the absorption and can be controlled over the relative delay between the two electromagnetic fields. We detected optical gain below 23.2 eV and above 23.5 eV. In these spectral regions the original pulse had no significant contributions. In perfect agreement with our theoretical predictions, we measure optical gain for negative and positive delays below and above the center energy of the excitation XUV pulse.

The behavior we observed is generic for any TLS and the underlying mechanism – control in time domain – offers a new and intuitive way for the creation and control of optical gain in this spectral region. We expect our universal model to be applicable in entirely different classes of systems ranging from molecules to low-dimensional solid-state systems in case of proper scaling of photon energies and intensities.

**Methods Summary**

We use a commercially available multi-pass, two-stage Ti:Sapphire amplifier (Femtolasers Femtopower V CEP) and focus the output (1 kHz, 1.2 mJ, 796 nm, 30 fs) into an Ar-filled gas cell to generate attosecond pulse trains (APTs) in the extreme ultraviolet (XUV). The use of shorter pulses for the high-order harmonic generation (HHG) is not possible in our case. It would result in spectrally broader harmonics populating several excited states and therefore preventing an off-resonant excitation. An aluminum filter with a thickness of 150 nm removes the residual IR radiation after the generation. Additionally, the filter compresses the APT in time due to its negative dispersion in this spectral region[27]. A temporal characterization of the APT with the RABITT (reconstruction of attosecond beating by interference of two-photon transitions[6,28]) technique gives us an average duration of a pulse in the APT of ~350 as. The duration of the isolated harmonic 15 can be estimated to about 12 fs, which was derived from a previous experiment[3]. The IR control pulse, which is separated from the main beam before the HHG, is recombined with the XUV pulse by a mirror with a center hole. A toroidal mirror focuses both beams, IR and XUV, into the pulsed interaction target. The target works with the same repetition rate as the laser and is synchronized to the arriving pulses. The interaction length in the target is 1 mm. This leads to an absorption of ~50% (corresponding to a particle density of $\sim 5 \times 10^{17}$ particles/cm$^3$)[3] of those harmonics that are located energetically above the ionization threshold of helium (24.59 eV). A spherical mirror that creates an astigmatic focus collects the transmitted XUV radiation. The resulting vertical line focus is well matched to the entrance slit of our grating spectrometer, where we detect the XUV radiation with a CCD camera. We filter out the IR control pulse with an additional aluminum filter between the interaction target and the spectrometer.

We change the delay between the XUV and infrared (IR) pulse by changing the length of the IR pulse beam path with a piezo-driven translation stage. An interferometric measurement with a helium-neon laser determined the stepsize of the delay to 107 as.

**Acknowledgments**

This work was supported by National Center of Competence in Research Molecular Ultrafast Science and Technology (NCCR MUST), research instrument of the Swiss National Science Foundation. Paula Rivière acknowledges a Juan de la Cierva contract grant from MICINN, and the COST action CM0702. We thank H. R. Reiss and M. Lucchini for fruitful discussions.

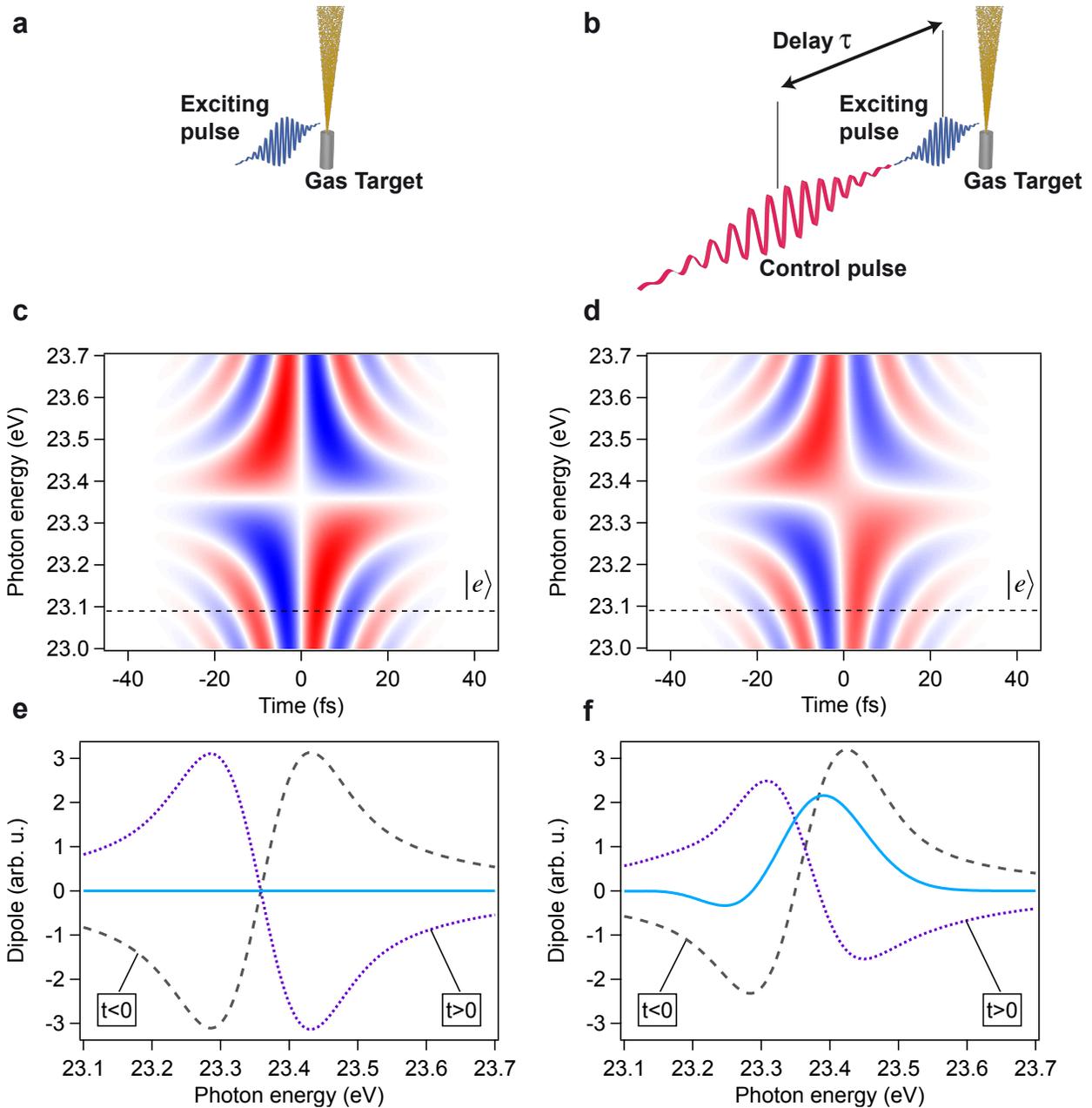

**Fig. 1. Time-frequency analysis of the dipole response of the two-level system for off-resonant excitation.**

(**a**) and (**b**) show the situation of excitation without the control pulse and with the control pulse delayed by $\tau$. (**c**) depicts the time-frequency representation of the dipole response. Red represents a positive and blue a negative response. In (**c**) and (**d**) the dashed line shows the energetic position of the excited state $|e\rangle$. (**e**) shows the time-integrated dipole response (TIDR) for negative times (dashed grey) and positive times (dotted purple). They differ only in the sign, which results in a zero TIDR when integrating over positive and negative times (solid light blue). (**d**) presents the dipole response with a loss term added to the model and a control pulse arriving 15 fs after the exciting pulse. The dipole response for positive times (dotted purple) in (**f**) is significantly reduced. Therefore, the temporal integration of the dipole response (solid light blue) is nonzero with positive and negative contributions.

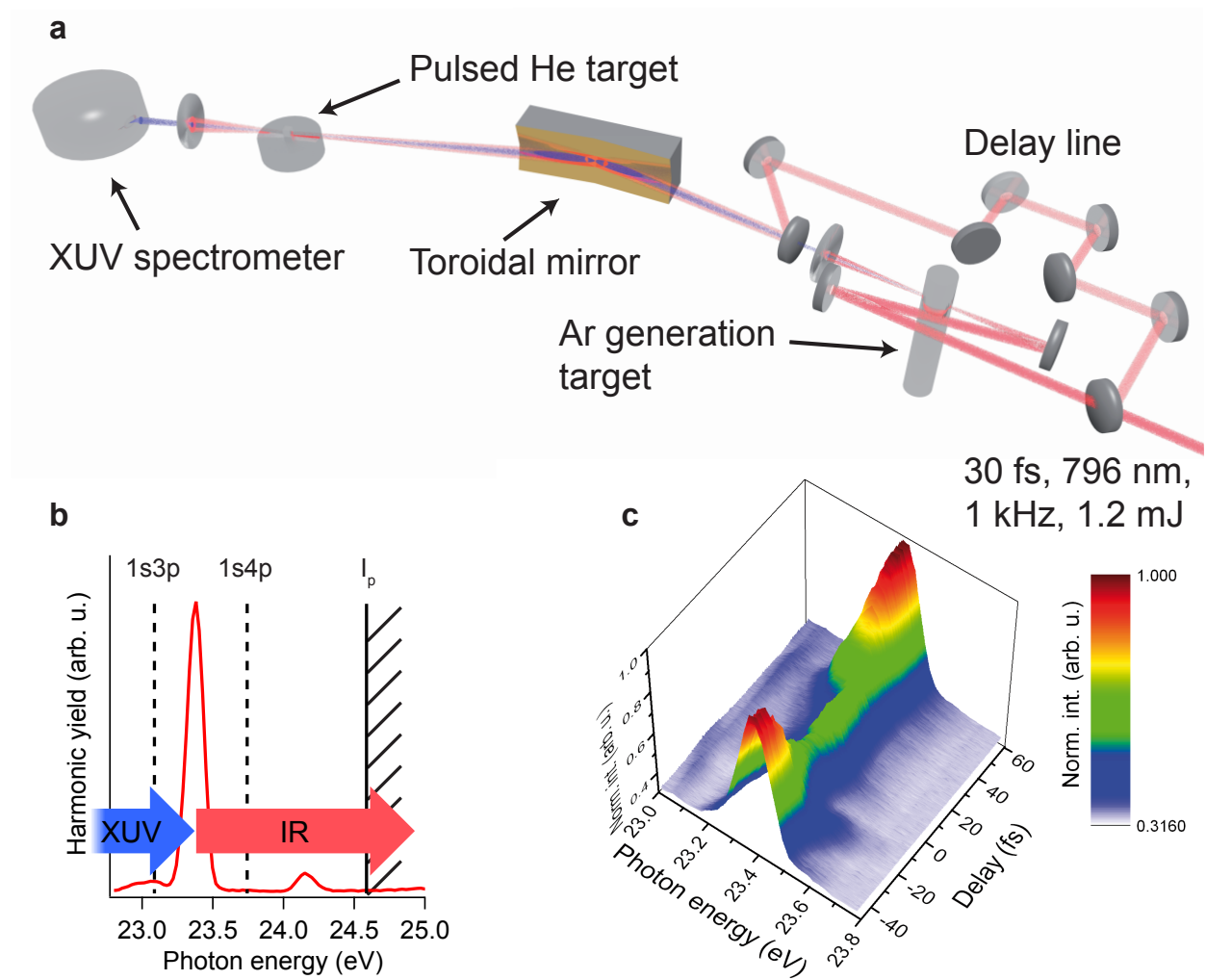

**Fig. 2. Setup of the experiment, investigated spectral window and transmitted XUV signal for different delays between the XUV and IR pulse.**

(**a**) An attosecond pulse train is generated by focusing 30-fs pulses of a Ti:Sapphire based laser system into a gas cell filled with argon. The IR control pulse is separated from the fundamental IR pulse before the HHG. After filtering out the residual IR light, the XUV pulse and the IR control pulse are collinearly recombined and focused into the He interaction target by a toroidal mirror. (**b**) The 15th harmonic is energetically located between the excited 1s3p and 1s4p states of He. Additionally the energy is below the ionization potential of He (24.59 eV). (**c**) The transmitted XUV signal shows strong absorption (about 50%) when both pulses overlap (~0 fs delay) due to a two-photon absorption (also shown in (**b**)).

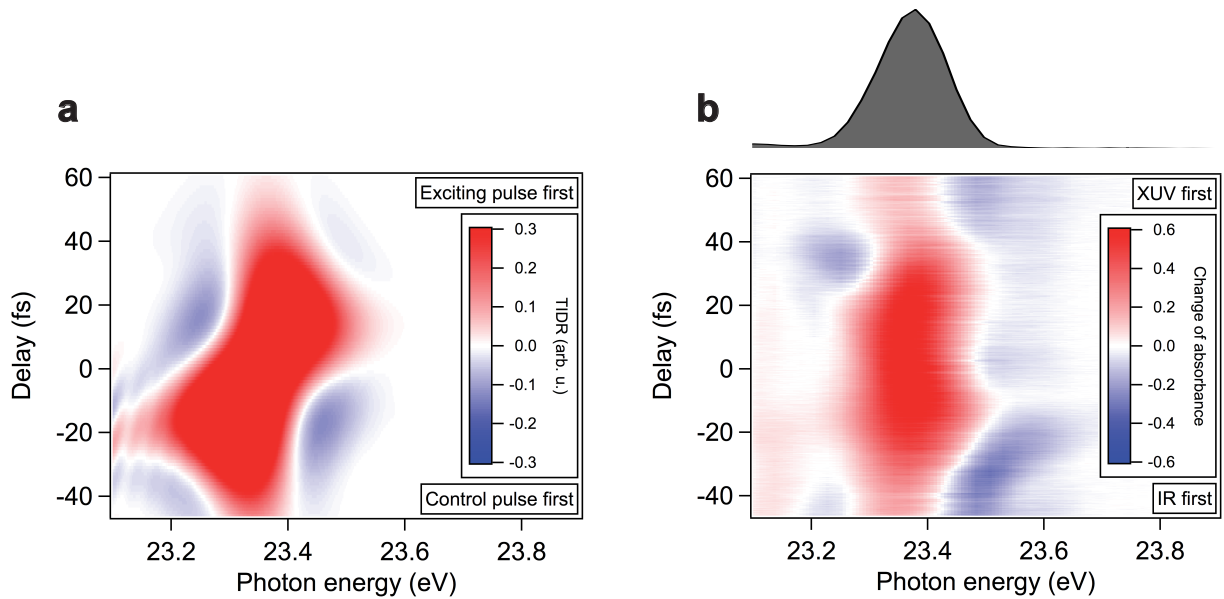

**Fig. 3. Calculated dipole response (a) and change of absorption (b) for different delays between the pulses.**

The change in optical density at XUV frequencies is plotted for different delays between the two pulses. For positive delays the XUV pulse is preceding the IR pulse. Red corresponds to a positive dipole response and is related to absorption, whereas blue stands for a negative dipole response (net emission, optical gain). On top of (**b**) the spectrum of the XUV pulse is shown. We also detect significant optical gain in spectral regions where the intensity of the original XUV pulse signal has already decreased to the noise level.